**Do We Need a Scientific Revolution?**

Nicholas Maxwell
(Emeritus Reader in Philosophy of Science at University College London)

**Abstract**
Many see modern science as having serious defects, intellectual, social, moral.  Few see this as having anything to do with the philosophy of science.  I argue that many diverse ills of modern science are a consequence of the fact that the scientific community has long accepted, and sought to implement, a bad philosophy of science, which I call standard empiricism.  This holds that the basic intellectual aim is truth, the basic method being impartial assessment of claims to knowledge with respect to evidence.  Standard empiricism is, however, untenable.  Furthermore, the attempt to put it into scientific practice has many damaging consequences for science.  The scientific community urgently needs to bring about a revolution in both the conception of science, and science itself.  It needs to be acknowledged that the actual aims of science make metaphysical, value and political assumptions and are, as a result, deeply problematic.  Science needs to try to improve its aims and methods as it proceeds.  Standard empiricism needs to be rejected, and the more rigorous philosophy of science of aim-oriented empiricism needs to be adopted and explicitly implemented in scientific practice instead.  The outcome would be the emergence of a new kind of science, of greater value in both intellectual and humanitarian terms.

I

Science suffers, in many different ways, from a bad philosophy of science.  This philosophy holds that the proper basic intellectual aim of science is to acquire knowledge of truth, the basic method being to assess claims to knowledge impartially with respect to evidence.  Considerations of simplicity, unity or explanatory power may legitimately influence choice of theory, but not in such a way that nature herself, or the phenomena, are presupposed to be simple, unified or comprehensible.  No permanent thesis about the world can be accepted as a part of scientific knowledge independent of evidence.  Furthermore, values have no role to play within the intellectual domain of science.  A basic humanitarian aim of science may be to help promote human welfare, but science seeks this by, in the first instance, pursuing the intellectual aim of acquiring knowledge in a way which is sharply dissociated from all consideration of human welfare and suffering.

    This view, which I shall call *standard empiricism* (SE) is generally taken for granted by the scientific community.  Scientists do what they can to ensure science conforms to the view.  As a result, it exercises a widespread influence over science itself.  It influences such things as the way aims and priorities of research are discussed and chosen, criteria for publication of scientific results, criteria for acceptance of results, the intellectual content of science, science education, the relationship between science and the public, science and other disciplines, even scientific careers, awards and prizes.[1]

II

    SE is, however, untenable, as the following simple argument demonstrates.  Physics only ever accepts theories that are (more or less) *unified*, even though endlessly many empirically more successful disunified rivals can always be concocted.  Such a theory, T (Newtonian theory, quantum theory, general relativity or the standard model), almost always faces some empirical difficulties, and is thus, on the face of it, refuted (by phenomena A).  There are phenomena, B, which come within the scope of the theory but which cannot be predicted

because the equations of the theory cannot (as yet) be solved. And there are other phenomena (C) that fall outside the scope of the theory altogether. We can now artificially concoct a disunified, "patchwork quilt" rival, T*, which asserts that everything occurs as T predicts except for phenomena A, B and C: here T* asserts, in a grossly ad hoc way, that the phenomena occur in accordance with empirically established laws, $L_A$, $L_B$ and $L_C$.

Even though T* is more successful empirically than T, it and all analogous rival theories are, quite correctly, ignored by physics because they are all horribly disunified. They postulate different laws for different phenomena, and are just assumed to be false. But this means physics makes a big, implicit assumption about the universe: it is such that all such "patchwork quilt" theories are false.

If physicists only ever accepted theories that postulate atoms even though empirically more successful rival theories are available that postulate other entities such as fields, it would surely be quite clear: physicists implicitly assume that the universe is such that all theories that postulate entities other than atoms are false. Just the same holds in connection with unified theories. That physicists only ever accept unified theories even though empirically more successful rival theories are available that are disunified means that physics implicitly assumes that the universe is such that all disunified theories are false.

But SE holds that no permanent thesis about the world can be accepted as a part of scientific knowledge independent of evidence (let alone against the evidence). That physics does accept permanently (if implicitly) that there is some kind of underlying unity in nature thus suffices to refute SE. SE is, in short, untenable.[2] Physics makes a big implicit assumption about the nature of the universe, upheld independently of empirical considerations - even, in a certain sense, in violation of such considerations: the universe possesses some kind of underlying dynamic unity, to the extent at least that it is such that all disunified physical theories are false. This is a secure tenet of scientific knowledge, to the extent that empirically successful theories that clash with it are not even considered for acceptance.

III

This substantial, influential but implicit assumption is however highly problematic. What exactly does the assumption amount to? What basis can there be for accepting it as a part of scientific knowledge?

In order to answer the first question, it is necessary to know how to distinguish unified from disunified physical theories. This has long been a fundamental unsolved problem in the philosophy of science. It is a problem in part because any theory can be formulated in many different ways, some unified, some highly disunified. Even Einstein recognized the problem but confessed he did not know how to solve it.[3]

The key to solving the problem is to attend, not to the theory itself, but to what it asserts about the universe, to its *content* in other words. A physical theory is unified if what it asserts - the content of the dynamical laws it specifies - are *precisely the same* throughout the range of possible phenomena to which the theory applies. A theory that specifies N different sets of laws for N ranges of possible phenomena, the laws of any one region being different from the laws of all the other regions, is disunified to degree N. For unity we require N = 1. This way of assessing the degree of unity of a theory is unaffected by changes of formulation. As long as different formulations all have the same content, the degree of unity will remain the same.

There is now a refinement. Sets of laws can differ in *different ways,* to *different extents*, in *more or less substantial ways*. Laws may differ in regions of space and time; or in ranges of other variables such as mass or relative velocity. Or a theory may, like the so-called standard model (the quantum field theory of fundamental forces and particles) postulate two or more different forces, or two or more kinds of fundamental particles (with different charges, masses or other properties).  Such a theory is disunified because in one range of possible phenomena to which the theory applies, one kind of force operates, or one kind of particle exists, and in another range a different force operates, or a different particle exists, there thus being different laws in different ranges of possible phenomena.  In addition to degrees of disunity there are, in short, different *kinds* of disunity, some more severe than others, depending on how different sets of laws are in different regions of phenomena.  Elsewhere I have argued that eight different kinds of disunity can be distinguished.[4]

The requirement that physics only accepts unified theories faces a further complication.  In some cases, presented with a theory disunified to degree $N = 3$, let us say, we can restore unity of theory in an entirely artificial way by splitting the one disunified theory into three unified theories.  In order to exclude this ruse, we need to formulate the requirement concerning unity in such a way that it applies to *all* fundamental dynamical theories (and to phenomenological laws when no theory exists).  Physicists in effect demand of an acceptable new fundamental theory that it is such that it *decreases* both the *kind* (i.e. the severity) and the *degree* of the disunity of the totality of fundamental physical theory when it replaces predecessor theories or laws.  A new theory must, in short, increase the unity of all fundamental physical theory, in addition to being sufficiently empirically successful, in order to be accepted as an addition to theoretical scientific knowledge.  Seriously disunified theories are not considered, whatever their empirical success might be, because they do not enhance overall theoretical unity.

It is this persistent, implicit demand for increased theoretical unity that commits physics to a persistent, substantial assumption about the nature of the universe.

IV

But what should this assumption be?  Should physics assume, boldly, that the universe is such that the yet-to-be-discovered true physical "theory of everything" is fully unified (in the sense explicated above)?  Or should physics assume, more modestly, that the universe is such that the true theory of everything is at least more unified than the current totality of fundamental physical theory (new, empirically successful but disunified theories being rejected because they clash with this assumption)?

Some such assumption must be made if the empirical method of science is to work at all - since otherwise physics would be drowned in an ocean of empirically successful but grossly disunified theories, and scientific progress would come to an end.  Whatever assumption is made, it is almost bound to be false.  We do not *know* that the universe is unified.  Even if it is, almost certainly it is not unified in the way current theoretical knowledge in physics suggests it is.

Contradictory considerations govern choice of assumption.  The more specific and substantial we make the assumption, the greater the help we will receive with developing new physical theories - as long as the assumption is correct.  On the other hand, the more specific and substantial the assumption is, the greater the chance, other things being equal, that it is false.

In order to resolve this dilemma, and give ourselves the best chances of learning, making progress, eliminating error, and improving our ideas, we need to see science as making, not one assumption, but a hierarchy of assumptions, these assumptions becoming less and less specific and substantial as one goes up the hierarchy, and thus more and more likely to be true, and more nearly such that their truth is required for science, or the pursuit of knowledge, to be possible at all: see figure 1.  At the top there is the thesis that the universe is such that we can acquire some knowledge of our local circumstances.  This is not an assumption we need ever reject since, if it is false, we cannot acquire knowledge whatever we assume.  As we descend the hierarchy, assumptions become increasingly substantial, increasingly likely to be false and in need of revision. At level 5 there is the thesis that the universe is comprehensible in some way or other, there being some one kind of explanation for everything that occurs. At level 4 there is the thesis that the universe is physically comprehensible - it being such, in other words, that the true theory of everything is unified. (To say that a physical theory is unified is equivalent to saying that it is explanatory and, if it is a theory of everything, that the universe it depicts is physically comprehensible.) At level 3 there is the thesis that the universe is physically comprehensible in some more or less specific way. Ideas, here, have changed dramatically over the centuries. Once there was the idea that everything is made up of corpuscles that interact by contact; then the idea that everything is made up of point-particles that interact by means of a force at a distance; then the idea that there is a unified field; nowadays there is the idea that everything is made up of quantum strings. At level 2 there is current accepted fundamental physical theory, at present the standard model and general relativity, and at level 1 there is the mass of established empirical data.

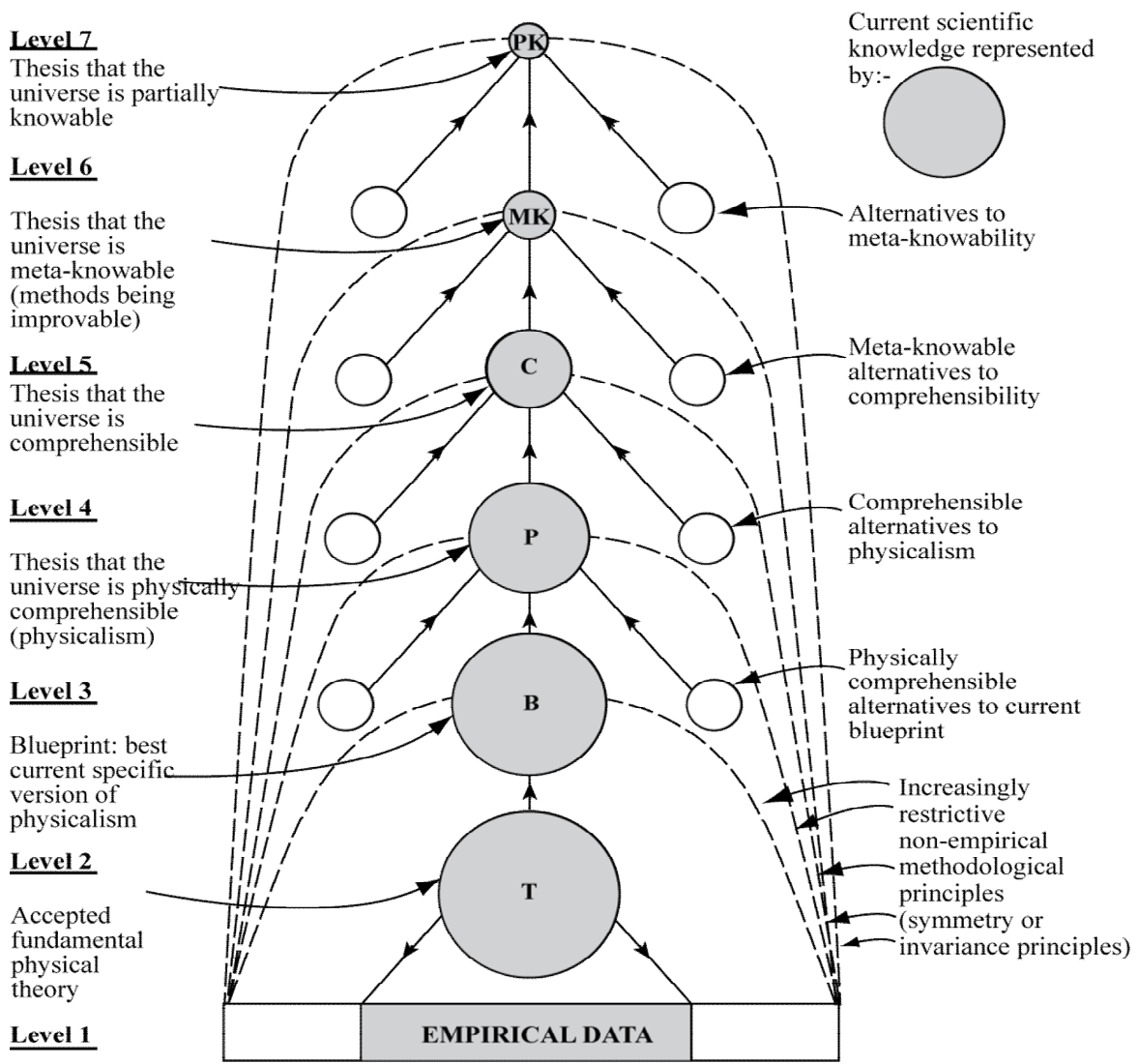

**Figure 1: Aim-Oriented Empiricism**

Associated with each thesis there is a methodological rule (represented by dotted lines in the diagram) which asserts: accept that thesis one down in the hierarchy which, as far as possible (a) is compatible with the thesis above and (b) best accords with, and best promotes, empirically successful theories at level 2.

The thesis at level 7 is almost certainly true, the currently accepted thesis at level 3 almost certainly false. As we descend the hierarchy, we move at some point from truth to falsity. The whole idea of the hierarchy is to concentrate criticism and revision where it is most likely to be needed, low down in the hierarchy. A framework of relatively stable, unproblematic assumptions and associated methods is created (high up in the hierarchy) within which much more specific, problematic assumptions and associated methods (low down in the hierarchy) can be critically assessed, revised and developed so as to give maximum help with the task of improving theoretical and empirical knowledge at levels 2 and 1. In short, according to this view, as we improve our empirical knowledge, we improve assumptions and associated methods at levels 3, and perhaps 4: we improve our knowledge-about-how-to-improve knowledge. There can be something like positive feedback between improving knowledge, and improving knowledge-about-how-to-improve-knowledge. Science adapts its nature to what it finds out about the universe.

Another way of putting the matter is to say that the basic intellectual *aim* of science is not truth *per se* (as standard empiricism holds) but rather truth *presupposed to be explanatory* - *explanatory* truth, in other words. Because this aim is profoundly problematic, it is important that it is represented in the form of a hierarchy of aims and associated methods - metaphysical assumptions implicit in these aims becoming increasingly insubstantial as one ascends the hierarchy, and thus increasingly likely to be true - a framework of relatively unproblematic aims and methods thus being created within which more specific and problematic aims and methods can be critically assessed and improved, as science proceeds. This way of putting the matter is important because it makes it possible to generalize scientific methodology, so conceived, so that it becomes fruitfully applicable to worthwhile human endeavours with problematic aims other than science, a point I will take up below.

Natural science puts something close to this hierarchical view into practice, but in a way that is constrained and handicapped by general allegiance to standard empiricism (SE), and it is this which damages science in a variety of ways, as we shall now see.

## V

How then does general acceptance and attempted implementation of SE damage science? How would acceptance and explicit implementation instead of the hierarchical view I have just outlined, which elsewhere I have called "aim-oriented empiricism" (AOE),[5] benefit science? Here are eight ways in which the move from SE to AOE would be beneficial.

1. AOE provides a more rigorous conception of science. An elementary requirement for rigour is that assumptions that are substantial, influential, problematic and implicit need to be made explicit so that they can be criticized, alternatives formulated and considered, in the hope of eliminating error and improving such assumptions. SE fails this requirement for rigour in failing to acknowledge the substantial, influential and problematic metaphysical (i.e. untestable) assumption that - at the very least - the universe is such that all disunified theories are false. AOE, by contrast, not only acknowledges such an assumption but, in addition, provides a framework within which what is most problematic can be subjected to severe, sustained criticism and attempted improvement, so as to help promote scientific progress. The hierarchy of assumptions of AOE might almost be construed as the outcome of repeated applications of the above requirement of rigour.

  A sign of the greater rigour of AOE over SE is provided by the fact that three fundamental problems in the philosophy of science, which cannot be solved granted SE, are solved within the framework of AOE. These are (1) the problem of what it means to say of a physical theory that it is unified (discussed above in section III), (2) the problem of what it can mean to hold that science makes progress if it proceeds from one false theory to another (the problem of verisimilitude) and, most serious of all, (3) the problem of induction. Elsewhere I have shown that these problems, unsolvable granted SE, can be solved within the framework of AOE.[6]

2. The greater rigour of AOE is no mere formal matter. It makes explicit the "positive feedback" feature of scientific method - the way in which methods for improving knowledge can themselves be improved, as science progresses, within a framework of persisting assumptions and meta-methods. Every scientist would agree that "positive feedback" of this type is an essential feature of scientific method at the empirical level. New knowledge leads to the development of new instruments, new experimental tools and techniques, which in turn may massively accelerate the acquisition of new knowledge. That something similar can go on at the theoretical level has not been properly acknowledged or understood, because of general acceptance of SE. For, whereas AOE stresses that methods (associated with lower level assumptions) improve as science progresses, SE specifies a fixed aim and fixed methods. Assumptions and associated methods of science have improved over the centuries (or we would still be stuck with pre-Galilean, Aristotelian science), but it has come about in an implicit, almost furtive fashion, retarded by general allegiance to SE.[7]

3. In stressing the vital, "positive feedback" interplay between methods and theories, AOE does far better justice to scientific practice than does SE. This aspect of AOE physics is especially apparent in Einstein's development of special and general relativity, and in the role of symmetry principles in modern physics.[8] Special relativity is a physical theory; it is a thesis about the nature of space and time (space-time is Minkowskian); and it is a methodological principle (an acceptable physical theory must be Lorentz invariant). AOE holds that methodological principles of this type may need to be revised or rejected - and just this happened, of course with the advent of general relativity (which asserts that space-time, in the presence of mass or energy, is curved and not flat). The principle of equivalence, associated with general relativity, has a status and role somewhat similar to that of Lorentz invariance of special relativity - as do gauge invariance and supersymmetry, associated with the standard model and string theory respectively.

4. So far AOE has been discussed as if relevant only to theoretical physics. But other branches of natural science have problematic aims too - in that aims make problematic assumptions about the world (assumptions usually taken from some more fundamental science). This is true, for example, of cosmology, chemistry, molecular biology, geology, ethology, neuroscience. Here too, in accordance with the basic idea of AOE, problematic aims need to be represented in the form of a hierarchy, aims becoming less problematic as one goes up the hierarchy, a framework of relatively unproblematic aims and methods being created in this way within which much more specific and problematic aims and associated methods may be criticized, alternatives being developed and assessed. The hierarchical, meta-methodological structure of AOE is relevant to all of natural science, and to all specialized disciplines within natural science, and not just to theoretical physics. SE, with its fixed aim for science, and its fixed methods, fails to do justice to any of this.[9]

5. In moving from SE to AOE there is a profound enhancement in the scope of scientific knowledge and understanding. Granted SE, scientific knowledge consists of (1) empirical results, and (2) accepted laws and theories. By contrast, granted AOE, scientific knowledge consists, in addition to (1) and (2), the level 4 thesis that (3) the universe is physically comprehensible (i.e. is such that the true theory of everything is unified) - a thesis I shall call *physicalism*.

Granted SE, physicalism cannot be a part of scientific knowledge because, being metaphysical, it can be neither verified nor falsified empirically, and most certainly has not predicted empirical phenomena, and thus achieved empirical success (which is what a theory must do, according to SE, if it is to become a part of scientific knowledge). Granted AOE, however, physicalism emerges as an especially secure part of theoretical scientific knowledge since all physical theories that clash with it too severely are rejected, or not even considered, however empirically successful they might be if considered.

All scientific knowledge is conjectural in character, as Karl Popper tirelessly argued.[10] This applies to physicalism too, of course. For all we can know for certain, physicalism may be false - as may be our best current theories of physics, the standard model and general relativity. (Physicalism implies, indeed, that these theories are false.) What arguments in support of AOE reveal, however, is that physicalism is a more secure part of (conjectural) theoretical scientific knowledge than our best current theories - and is certainly as much a part of current knowledge as these theories.[11]

According to AOE, then, and in sharp contrast to SE, science already provides us with (conjectural) knowledge about the ultimate nature of the universe: it is physically comprehensible. Some kind of physical entity, some sort of field, exists everywhere, unchanging, and determines (perhaps probabilistically) the way everything that changes does change.

This represents, not just an enormous increase in the scope, the content, of scientific knowledge; it is an increase that is of profound significance for humanity. Physicalism threatens the value of human life. If it is true, how can there be free will? How can there be consciousness? What becomes of everything that physicalism seems to leave out, the whole world of human experience, meaning and value? The transition from SE to AOE serves to highlight just how fundamental and intellectually urgent these problems are (in that they are engendered by scientific knowledge, and not merely by speculative metaphysics).[12]

6. AOE places far greater emphasis on the importance of the search for explanation and understanding in science than does SE. For a new theory to be accepted, granted SE, what really matters is that the theory is empirically successful. Considerations of simplicity, unity, explanatory character, may play a role as well, but as SE fails to explicate clearly what these non-empirical considerations are, and why they should be relevant, their influence is not nearly as important as empirical considerations. The move to AOE changes this situation dramatically. What the demand for unity means becomes quite clear. Why it is a legitimate, rational demand is also clear. And this demand for unity is so central to science that it persistently over-rides empirical considerations, as we have seen. This happens all the time in scientific practice, in defiance of SE, when empirically successful, disunified theories are ignored. And AOE provides a rationale for permitting unity or explanatoriness to over-ride empirical considerations. The quest to explain and understand becomes central to science, granted AOE, in a way which is not the case, granted SE.

That AOE does far better justice to the search for explanation and understanding in science than SE is strikingly borne out by the case of orthodox quantum theory (OQT). Those who developed OQT (Heisenberg, Bohr, Born and others) despaired of solving the quantum wave/particle dilemma (required for understanding) and, as a result, developed OQT, not as a theory about quantum systems per se, but rather as a theory about the results of performing measurements on such systems. The extraordinary empirical success of OQT led to its general acceptance, even though it fails to provide real explanation and understanding of quantum phenomena.

Viewed from the perspective of AOE, all this looks very different. Because OQT is a theory about the results of performing measurements on quantum systems (and not a theory about quantum systems per se), OQT is composed of a quantum part, and some part of classical physics for a treatment of measurement. OQT is, in other words, unacceptably *disunified* (being made of incompatible components). If AOE had been generally accepted when OQT was being developed, OQT might have been tolerated as an empirically successful theoretical scheme, but it would not have been regarded as an acceptable theory, precisely because of its gross disunity, its failure to provide explanation and understanding of the quantum domain. It is no accident, incidentally, that Einstein vehemently opposed OQT; he held a view close to AOE.

General acceptance of AOE in the 1920s would have led to recognition of the scientific importance of curing the unity defects of OQT, and developing a more acceptable version of quantum theory. As a result, we might today have such a version of the theory - something we still do not have, eighty years later! Here is a graphic example of the way acceptance of SE can damage the content of science.[13]

7. Granted SE, there is no such thing as a rational method for the discovery of fundamental new theories. The only guidelines available are existing theories, but new theories almost invariably contradict predecessor theories. By contrast, AOE does provide science with a rational method of discovery - even though fallible and non-mechanical. In order to develop new fundamental physical theories (the hardest kind of case to consider), physicists need to resolve clashes between existing fundamental theories, and between these theories and metaphysical ideas at levels 3, and 4. Something like this method was employed by Einstein in discovering special and general relativity, and somewhat similar methods were employed by Yang, Mills, Weinberg, Salam, Gell-Mann and others in discovering elements of the standard model.[14] As I have already remarked, science puts something like AOE into practice, but in an awkward, furtive, retarded way, handicapped by allegiance to SE.

8. The transition from SE to AOE would have fruitful implications for science education, and for public understanding and appreciation of science. The intellectual content of science, shaped by allegiance to SE, consists of theory and evidence. Metaphysics, philosophy, epistemology, questions about the meaning and value of human life must all be ruthlessly excluded from science, according to SE, to preserve its rationality, its intellectual integrity. The intellectual content of science, understood in this SE way, tends to be technical, esoteric, unintelligible and unappealing to many pupils, students and members of the public. There is the danger that science is reduced to being merely the highly technical enterprise of predicting more and more phenomena more and more accurately.

Science would be very different if shaped by allegiance to AOE. The metaphysical thesis of physicalism would be acknowledged to be a central, fundamental tenet of scientific knowledge. This is a thesis everyone can understand, unlike scientific theories like the standard model or general relativity. Methodological, epistemological and philosophical questions would be acknowledged to be an integral part of AOE science. Science becomes much more like natural philosophy, what it was for Galileo and his successors, before it became malformed by SE. The quest to understand becomes much more important, as we have seen. Science education (whether involving children, students or members of the public) would need to include discussion of ultimate scientific/ philosophical questions that we can all understand and appreciate. What kind of universe is this? Does science really tell us that physicalism is true? If it is true, what becomes of consciousness, free will, the meaning and value of human life? Instead of being primarily the technical, esoteric, unintelligible affair engendered by SE, AOE science becomes a dramatic, exhilarating and even alarming quest and adventure, "the greatest spiritual adventure of mankind" as Popper has called it, open to all, intelligible at the most fundamental level to all. Science would become again what it should be, somewhat like music, technical and professional in some respects but fundamentally open for all to enjoy and participate in, a vital part of culture and not something shielded from it.[15]

## VI

In the next two sections I discuss two generalizations of AOE. The first, discussed in this section, has to do with AOE applied to science. The second, discussed in the next section, has to do with AOE applied to all worthwhile endeavours with problematic aims other than science.

So far I have argued that the basic aim of science is not truth *per se*, but rather *explanatory* truth. But this latter is a part of a more general aim of science of seeking *valuable* truth - of value either because of its intrinsic or intellectual value (of value because it enables us to explain and understand or because of its inherent interest), or because it enables us to achieve other goals of value - health, communications, travel, prosperity, etc. - by means of technology, or in other ways.

In order for a result to be accepted for publication in a scientific journal, let alone accepted as a part of scientific knowledge, it is not enough that the result be new and sufficiently well established. It must, in addition, be judged to be of sufficient interest, significance or importance. Values, of one kind or another, thus play a decisive role in deciding what enters, and what is excluded from, the body of scientific knowledge. A science which accumulated a vast store of knowledge about facts all irredeemably trivial and useless would not be judged to be making splendid progress; it would, quite correctly, be held to be stagnant and decadent.

It is *desirable* that science should seek valuable truth. And that values, of one kind or another, should influence the aims of research is *inevitable*. Infinitely many facts about the world are all around us, awaiting potential scientific investigation: inevitably decisions about what is *important* will influence what is, and what is not, studied.

Values influence the aims and content of science in a way quite different from the influence of metaphysical assumptions, discussed above in sections II to IV. Values do not, or ought not to, influence decisions about truth and falsity; rather, inevitably and quite properly, as I have said, they influence decisions about whether a result is sufficiently significant to enter the body of scientific knowledge, or even be published.

If metaphysical assumptions implicit in the aim of seeking explanatory truth are problematic, then it must be said that value assumptions, inherent in the aim of seeking valuable truth are, if anything, even more problematic. Of value to whom? When? In what way? How is one to decide between the very different values of science pursued for its own sake - for example, for the sake of explanation and understanding - and science pursued for the sake of achieving other aims of value - health, prosperity, etc.?

The argument here is exactly the same as before. If science is to stand any chance of pursuing aims that are both scientifically achievable and of value to achieve it is vital that possible and actual, highly problematic, aims be explicitly articulated as an integral part of science, so that they can be criticized, alternatives being developed and considered. Conjectures about (1) what is scientifically discoverable, and (2) what it would be of value to discover, need to be articulated and scrutinized in an attempt to discovery that highly problematic region of overlap of (1) and (2), the scientifically discoverable that is genuinely of value. Such conjectures concerning actual and possible aims need to be discussed in journals and forums open to scientists and non-scientists alike.

All this is encouraged by the generalized version of AOE which recognizes that a basic aim of science is valuable truth. SE, however, holds that the aim of science is truth as such, and denies that values have any role to play within the intellectual domain of science. Instead of recognizing the vital need to have three domains of scientific discussion, (1) evidence, (2) theory, and (3) aims, SE recognizes only the first two. And as a result of the profound influence that SE has long exercised over science, institutional means for the sustained imaginative and critical discussion of research aims and priorities have not been developed. Such matters are decided by grant giving bodies, committees, individual scientists, leaders of research groups.[16]

The outcome that one would expect of this SE failure to promote sustained discussion of problematic aims is that the priorities of research come to reflect, not the noblest and best interests of humanity - such as help for the poor and those who suffer, the search for understanding - but rather the interests of those who pay for science, the wealthy and powerful, and the interests of scientists themselves, including such non-intellectual matters as careers and status.

Such is indeed the case. Much scientific and technological research is devoted to the interests of wealthy countries and not to the interests of the billion or so who live in abject poverty. Medical research is devoted primarily to the diseases of the wealthy, not the poor. And there is the scandal of military research. In the UK, 30% of the budget for research and development is spent on the military. In the USA it is 50%.[17] In our world, fraught with gross inequalities, injustices, conflict and war, one may well wonder whether this expenditure is in the best interests of humanity. Striking, too, is the general silence about the matter, the failure of the scientific community to speak up about it. Whereas AOE insists that it is the professional duty of the scientific community to discuss and publicize these matters, SE implies that this lies beyond the scientist's brief, which has to do, exclusively, with the acquisition of value-neutral knowledge.

The problematic aims of science require further elaboration. The aim of seeking valuable truth needs to be regarded as a means to the realization of a farther social, or humanitarian aim: to make knowledge of valuable truth available to be used by people in their lives to enrich the quality of life, either culturally or intellectually, by enhancing personal knowledge and understanding of aspects of the world, or practically, to achieve other goals of value (health, prosperity, etc.). Inevitably scientific results are used to transform society. Science ought to do what it can to ensure that results are used for the benefit of all, and not in ways which are harmful. To this extent, science has a moral, a social, even a political, goal.

But this political or humanitarian goal is, if anything, even more problematic than the aims already discussed. Here, as before, it is vital that science promotes open, imaginative and critical discussion of actual and possible human uses of science, and relevant political policies and programmes, in an attempt to ensure that uses and policies of genuine benefit to humanity will be adopted. This is encouraged, indeed demanded, by AOE, but SE holds that it has nothing to do with science whatsoever.

Once again, because of the influence of SE, science has not developed institutional means for open, sustained, imaginative and critical discussion of the humanitarian aim of science to help enhance the quality of human life (by intellectual, technological and educational means). And as a result, as is to be expected, the impact of science on society has not always been of the best.

Modern science has, of course, been of immense benefit to humanity in countless ways. The modern world is inconceivable without it. But there is an underside to the blessings of science. They are not, to begin with, equally distributed throughout the world. Some billion people, as I have already mentioned, live in abject poverty, not much benefited by science. Science has made possible rapid population growth, modern agriculture and industry which in turn have led to pollution of earth, sea and air, destruction of natural habitats such as tropical rain forests, and rapid extinction of species. Science has made it possible for modern warfare to acquire its lethal character, and has led to the threats posed by modern armaments, conventional, chemical, biological and nuclear. Science has even played a role in engendering the AIDS crisis (AIDS being spread by modern methods of travel). And over everything hangs the menace of global warming, with its attendant threats of drought, storms floods, and death, threatening to intensify other crises (global warming being the outcome of population growth, modern industry and travel, all made possible by science).

AOE, recognizing the profoundly problematic character of the humanitarian scientific aim of helping to enhance the quality of human life, would anticipate problems such as these, and would require scientists and non-scientists alike to develop both scientific research and political policies designed to alleviate them. SE, by contrast, places all this outside the domain of science, and makes it no part of the professional task of the scientist to come to grips with such issues, as scientist. General acceptance of SE has thus played a role in allowing these crises to develop.

An important part of the humanitarian aim of science is the intellectual or educational one of enabling non-scientist to use science so as to enhance their personal knowledge and understanding of, and curiosity about, the world around them. Here too science must be judged to be only partially successful. Billions of people alive today are ignorant of even the most elementary aspects of the scientific picture of the world, and may well have a hostile attitude towards science. Acceptance of SE, rather than AOE, has contributed to this failure, partly in excluding the scientific picture of the world as an item of scientific knowledge, partly
in down-playing the importance of public involvement with science by means of discussion and debate.[18]

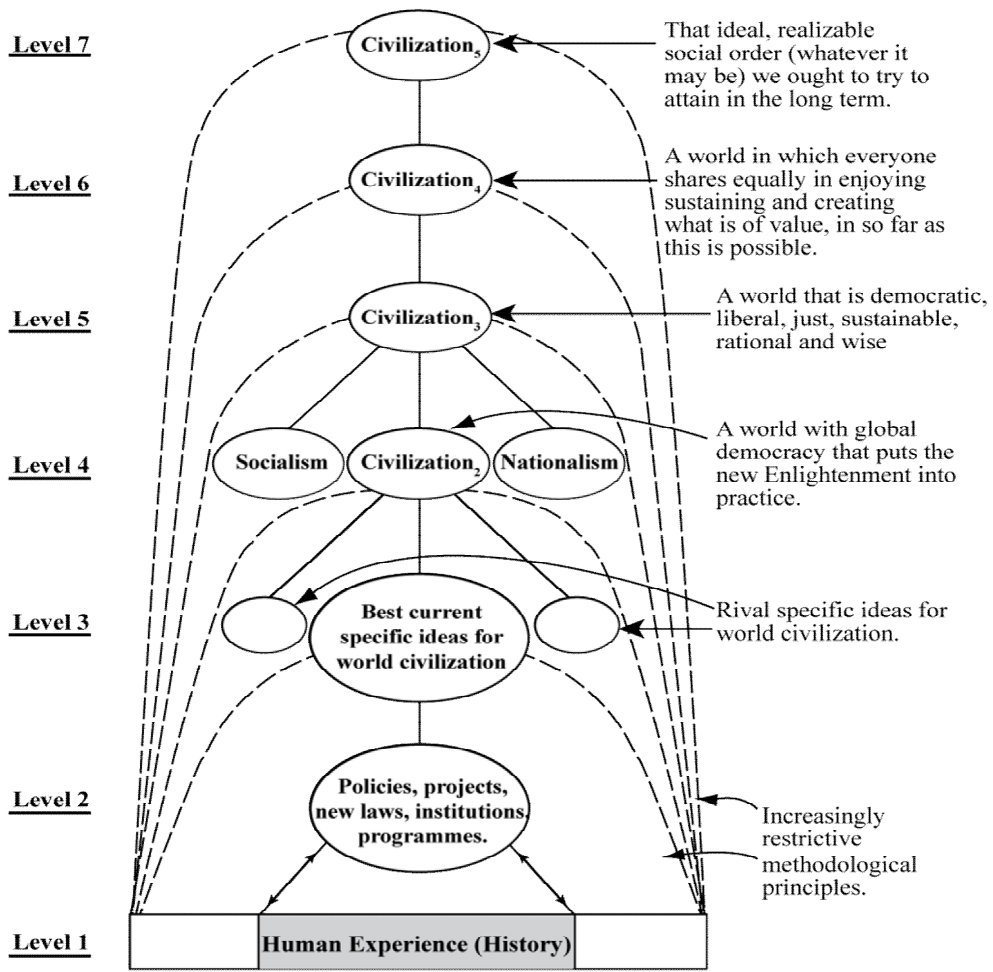

**Figure 2: Implementing Generalization of Aim-Oriented Empiricism in Pursuit of Civilization**

## VII

Science is of value in three ways: (1) intellectually, in enhancing knowledge, understanding and curiosity, (2) practically, via technology and in other ways, and (3) methodologically, as a quite extraordinarily successful example of learning, of making progress, which may well have fruitful implications for all sorts of other worthwhile human endeavours that struggle to meet with success, and make progress. All three are deeply problematic. (1) and (2) are, of course, well known, but (3) is nowadays almost universally overlooked, ignored and unused.

(3) involves generalizing the progress-achieving methods of science so that they become fruitfully applicable to other worthwhile pursuits. But to do this successfully it is vital to adopt AOE, and not SE, as one's conception of the progress-achieving methods of science. For it is not just in science that aims are problematic; in life, too, aims can be profoundly problematic. This is especially true of the humanitarian, political aim to create a better world. SE, generalized so as to apply to this and other pursuits, provides no help with improving problematic aims. AOE, generalized, by contrast, is specifically designed to help us improve problematic aims as we act. General acceptance of SE has had the effect of crippling this third, methodological use of science. This may well be the most damaging consequence of the failure of the scientific community to adopt AOE and repudiate SE.

The basic idea of (3) goes back at least to the French Enlightenment. The *philosophes*, Voltaire, Diderot, Condorcet and others, had the idea of learning from scientific progress how to make social progress towards an enlightened world. But in developing this idea, they blundered, and it was, unfortunately, their botched version of the idea that was taken up and subsequently built into the institutional structure of academia. It is from this that we still suffer today.

In order to put (3) - the Enlightenment idea - successfully into practice, three steps need to be got right:-
(a) The progress-achieving methods of science need to be correctly characterized.
(b) These methods need to be correctly generalized, so that they become fruitfully applicable to worthwhile human pursuits other than science (especially to those with problematic aims) - government, the pursuit of justice, prosperity, security, art, happiness, love.
(c) These generalized, progress-achieving methods then need to be embedded into the fabric of society and our lives, into institutions associated with government, the law, the economy, etc., and above all into efforts to create a just, peaceful, sustainable, democratic, liberal, prosperous, civilized world.

The *philosophes* got all three steps, (a), (b) and (c), wrong. They (a) took inductivism, a crude version of SE, for granted, which (c) they applied, not to *social life*, but rather to the task of creating *social science*. This was developed throughout the 19th century, and built into academia in the early 20th century with the creation of departments of economics, anthropology, sociology, political science, psychology. The outcome is what we still have today, academia devoted primarily to the pursuit of knowledge.

But all this is a damaging mistake. Step (a) requires that we adopt AOE, not SE. Step (b) requires that AOE is generalized: whenever worthwhile but problematic aims are pursued, a hierarchy of aims (and associated methods) needs to be created, aims becoming increasingly unspecific and unproblematic as one goes up the hierarchy, in this way a framework of relatively unproblematic aims and methods being developed within which much more specific, problematic and controversial aims and methods can be improved as we act. Step (c) requires that this hierarchical meta-methodology be adopted and implemented by social life, by institutions other than science - especially by those whose basic aims are problematic. The aim of creating a better world is, for all sorts of reasons, profoundly problematic. Here, above all, the generalized version of AOE must be adopted and implemented. (See figure 2 for a cartoon version of what is required.)

Social inquiry, on this view, is not, in the first instance, social *science*, or the pursuit of *knowledge* at all; rather it is social *methodology*, or social *philosophy*, concerned to help humanity tackle its immense problems of living in more cooperatively rational ways than at present, and seeking to build into social life progress-achieving methods arrived at by generalizing AOE - the progress-achieving methods of natural science.

If the basic Enlightenment idea had been properly developed in this way, in the 18th and 19th centuries, we might have learned how to avoid some of the horrors of the 20th century, and some of the crises that now beset us, partly as a result of the pursuit of SE science. For humanity would have had in its hands (what we still do not have today) institutions of inquiry rationally designed and devoted to help us make progress towards a genuinely civilized world.[19]

VIII

Many see modern science as having serious defects, intellectual, social, moral. Few see this as having anything to do with the philosophy of science. I have argued that many diverse ills of modern science are a consequence of the fact that the scientific community has long accepted, and sought to implement, a bad philosophy of science, namely standard empiricism. The scientific community urgently needs to bring about a revolution in both the conception of science, and science itself. Standard empiricism needs to be rejected, and the more rigorous philosophy of science of aim-oriented empiricism needs to be adopted and explicitly implemented in scientific practice instead. The outcome would be the emergence of a new kind of science, of greater value in both intellectual and humanitarian terms.

Val di Bottoli, Tuscany, July 2008

**Notes**

[1] For a more detailed exposition of SE, see Maxwell (2007), pp. 32-51.  For grounds for holding scientists do, by and large, accept SE, see Maxwell (1998, pp. 38-45; 2007, pp. 145-156; 2004, pp. 5-6, note 5).

[2] For more detailed refutations of SE, see Maxwell (1998, ch. 2; 2004, ch.1; 2005; 2007, ch. 9).

[3] Einstein (1982, pp. 21-25).

[4] Maxwell (1998, pp. 89-93; 2004, appendix, section 2; 2007, ch. 14, section 2).

[5] A version of AOE was first expounded and defended in Maxwell (1974).  For subsequent elaborations see Maxwell (1976a; 1984; 1998; 2004; 2005; 2007, ch. 14).

[6] Maxwell (2007), ch. 14.

[7] According to AOE, the philosophy of science, conceived of as being about the aims and methods of science, has an important, influential role to play within science, as an integral part of science itself.  Most contemporary academic philosophy of science unfortunately takes versions of SE for granted and thereby condemns itself to being worse than useless.  According to SE, science has a fixed aim and fixed methods: the philosophy of science cannot itself be a part of science since it does not consist of empirically testable ideas.  SE philosophy of science, instead of performing the useful task of demolishing SE and arguing for its replacement, tries to justify SE, and of course fails.  Furthermore, SE philosophy of science is obliged to interpret itself as a meta-discipline, seeking to clarify what the aims and methods of science are, but not in any way affecting science itself.  In thus dissociating the philosophy of science from science itself, the discipline serves to undermine the very thing it claims to be seeking to understand, namely the rationality of science.  This requires, as we have seen, that the philosophy of science is an integral, influential part of science itself, and

not a distinct meta-discipline.

[8] See Maxwell (1993; 1998, pp. 123-140).

[9] See Maxwell (2004), pp. 41-47.

[10] Popper (1959; 1963).

[11] See Maxwell (1998, ch. 5; 2007, ch. 14, section 6).

[12] For my own attempt at solving these fundamental problems see Maxwell (1984, ch. 10; 2001).

[13] Over many years I have attempted to develop a version of quantum theory that solves the wave/particle problem and is testably distinct from OQT: see Maxwell (1972; 1976b; 1982; 1994; 1998, ch. 7; 2008a). The key idea is that the quantum domain is fundamentally *probabilistic*, the condition for probabilistic transitions to occur being that new particles, bound systems or stationary states are created as a result of inelastic collisions. Somewhat similar ideas, different in detail, have been put forward by Ghirardi *et al* (1986) and Penrose (1986).

[14] See Maxwell (1993; 1998, pp. 123-140 and 219-223; 2004, pp. 34-39.)

[15] I have sought to get this idea across in Maxwell (1976a; 2004, pp. 47-51; 2008b.)

[16] Recently, attempts have been made to create institutional means for the discussion of aims to which non-scientists can contribute, for example by the Royal Society, and by the E.S.R.C. Science in Society Programme, in the UK.

[17] See Langley (2005).

[18] The argument of this section is spelled out in much greater detail in Maxwell (1976a; 1984; 2004; 2007; 2008b). I should add that in recent years natural science has moved somewhat in the direction I have argued for (but independently of my work), as a result of growing awareness by scientists of environmental problems, especially global warming: see Maxwell (2007, chs. 11 and 12).

[19] The argument sketched in this section was first spelled out in some detail in Maxwell (1976a). A much fuller exposition is to be found in Maxwell (1984; see also 2004; 2007). More information about my work is available at www.nick-maxwell.demon.co.uk.